\documentclass[12pt, a4paper, twoside]{scrartcl}
\pdfoutput=1

\usepackage[tracking=true,expansion=true,stretch=15,shrink=15]{microtype}
\DeclareMicrotypeSet*[tracking]{my}
{ font = */*/*/sc/* } 
\SetTracking{encoding = *, shape = sc }{ 45 }
\SetProtrusion{encoding={*},family={bch},series={*},size={6,7}}
{1={ ,750},2={ ,500},3={ ,500},4={ ,500},5={ ,500},
	6={ ,500},7={ ,600},8={ ,500},9={ ,500},0={ ,500}}
\SetExtraKerning[unit=space]
{encoding={*}, family={qhv}, series={b}, size={large,Large}}
{1={-200,-200}, 
	\textendash={400,400}}

\usepackage{setspace}
\usepackage[
	left=1in,
	right=1in,
	top=0.7in,
	bottom=0.7in,
	includeheadfoot
]{geometry}
\usepackage[automark, headsepline]{scrlayer-scrpage}

\usepackage[]{todonotes}
\usepackage{lineno}
\usepackage{amssymb}
\usepackage{amsmath}
\usepackage{amsthm}
\usepackage{algorithm}
\usepackage{dsfont}
\usepackage{bm}
\usepackage{subfig}
\usepackage{enumerate}
\usepackage{graphicx}
\usepackage{float} 
\usepackage{tikz}
\usetikzlibrary{shapes,arrows,decorations.pathmorphing,graphs,positioning,backgrounds}
\usepackage[square,comma, sort, numbers]{natbib}
\usepackage{hyperref}
\usepackage{nameref}
\definecolor{TUMblue}{cmyk}{1, .54, .04, .19}
\hypersetup{
	colorlinks   = true, %Colours links instead of ugly boxes
	urlcolor     = TUMblue, %Colour for external hyperlinks
	linkcolor    = TUMblue, %Colour of internal links
	citecolor   = TUMblue %Colour of citations
}

\newtheoremstyle{new}{12pt}{12pt}{\itshape}{}{\bfseries}{.}{1em}{}
\theoremstyle{new}

\newcommand{\ind}{\mathds{1}}
\newcommand{\R}{\mathds{R}}

\newcommand{\wh}{\widehat}

\newcommand{\etae}{{\eta_e}}
\newcommand{\etah}{{\wh{\bm \eta}}}
\newcommand{\etahe}{{\wh{\eta}_e}}
\newcommand{\kmax}{q_{\mathrm{max}}}
\newcommand{\etab}{{\bm \eta}}

\allowdisplaybreaks
\newcommand*\patchAmsMathEnvironmentForLineno[1]{%
  \expandafter\let\csname old#1\expandafter\endcsname\csname #1\endcsname
  \expandafter\let\csname oldend#1\expandafter\endcsname\csname end#1\endcsname
  \renewenvironment{#1}%
     {\linenomath\csname old#1\endcsname}%
     {\csname oldend#1\endcsname\endlinenomath}}% 
\newcommand*\patchBothAmsMathEnvironmentsForLineno[1]{%
  \patchAmsMathEnvironmentForLineno{#1}%
  \patchAmsMathEnvironmentForLineno{#1*}}%
\AtBeginDocument{%
\patchBothAmsMathEnvironmentsForLineno{equation}%
\patchBothAmsMathEnvironmentsForLineno{align}%
\patchBothAmsMathEnvironmentsForLineno{flalign}%
\patchBothAmsMathEnvironmentsForLineno{alignat}%
\patchBothAmsMathEnvironmentsForLineno{gather}%
\patchBothAmsMathEnvironmentsForLineno{multline}%
}

\begin{document}
	\title{Model selection in sparse high-dimensional vine copula models with application to portfolio risk}

\author{Thomas Nagler\footnote{Corresponding author, Department of Mathematics, Technische Universit{\"a}t M{\"u}nchen, Boltzmanstra{\ss}e 3,  85748 Garching, Germany (email: \href{mailto:thomas.nagler@tum.de}{thomas.nagler@tum.de})}\, , 
Christian Bumann\footnote{Department of Mathematics, Technische Universit{\"a}t M{\"u}nchen, Boltzmanstra{\ss}e 3,  85748 Garching, Germany}, \,
Claudia Czado\footnote{Department of Mathematics, Technische Universit{\"a}t M{\"u}nchen, Boltzmanstra{\ss}e 3,  85748 Garching, Germany (email: \href{mailto:cczado@ma.tum.de}{cczado@ma.tum.de})}}

\date{\hspace{3pt} \normalsize\today}

\maketitle

%---------------------------------------------------------%

\begin{abstract} 
\noindent {\bfseries \sffamily Abstract}\\
Vine copulas allow to build flexible dependence models for an arbitrary number of variables using only bivariate building blocks. The number of parameters in a vine copula model increases quadratically with the dimension, which poses new challenges in high-dimensional applications. To alleviate the computational burden and risk of overfitting, we propose a modified Bayesian information criterion (BIC) tailored to sparse vine copula models. We show that the criterion can consistently distinguish between the true and alternative models under less stringent conditions than the classical BIC. The new criterion can be used to select the hyper-parameters of sparse model classes, such as truncated and thresholded vine copulas. We propose a computationally efficient implementation and illustrate the benefits of the new concepts with a case study where we model  the dependence in a large stock stock portfolio.  
\end{abstract}

%---------------------------------------------------------%

\pagestyle{scrheadings}
\clearscrheadings
\lohead{Model selection in sparse high-dimensional vine copula models}
\rohead{\pagemark}
\lehead{T.\ Nagler, C.\ Bumann, C.\ Czado}
% * <t.nagler@gmx.net> 2018-01-16T16:12:02.948Z:
%
% ^.
\rehead{\pagemark}
	\section{Introduction}

Following the 2008 financial crisis, academic research and public media identified unrealistic mathematical models for the dependence as one of its key causes \citep{donnelly2010, salmon2009}. In the aftermath, modeling the dependence between financial assets became a hot topic in finance and statistics. One of most promising tools that emerged are vine copulas \citep{bedford2001,aas2009}. Vine copula models build the dependence structure from bivariate building blocks, called \emph{pair-copulas}. Each pair-copula captures the conditional dependence between a pair of variables. Because each pair-copula can be parametrized differently, vine copulas allow each pair to have a different strength and type of dependence. This flexibility is the key reason why vine copulas became so popular for modeling dependence between financial assets; for a review, see \citep{aas2016}.

The flexibility of vine copulas comes at the cost of a large number of model parameters. A vine copula on $d$ variables consists of $d(d-1)/2$ pair-copulas, and each pair-copula can have multiple parameters. In financial applications, the number of model parameters quickly exceeds the number of observations. Suppose each pair-copula can have up to two parameters. Then a model for 50 assets has $2\,450$ possible parameters, a model for 200 has almost $20\,000$. On the other hand, five years of daily stock returns consist of roughly $1\,250$ observations.

In such situations, there are two major challenges that we want to address: the risk of overfitting and the computational burden to fit thousands of parameters. Both issues make it desirable to keep the model sparse. A vine copula model is called \emph{sparse} when a large number of pair-copulas are the independence copula. The key question is now which of the pair-copulas we should select as independence copula. 

The most common strategy is to select the pair-copula family by either AIC or BIC \citep{akaike1974, schwarz1978}. The AIC is known to have a tendency of selecting models that are too large, contradicting our preference for sparsity (see \citep{claeskens2008model}). The BIC on the other hand is able to consistently select the true model, but only when the number of possible parameters grows sufficiently slowly with the sample size $n$. As explained above, this assumption should be seen critically in a high-dimensional context. Another unpleasantry is that BIC is derived under the assumption that all models are equally likely. Under this assumption, we expect only half of the pair-copulas to be independence, a model we would not consider sparse. 

We propose a new criterion called  \emph{modified Bayesian Information Criterion for Vines} (mBICV) that addresses both issues and is specifically tailored to vine copula models. It grounds on a modification of the prior probabilities that concentrates on sparse models and is motivated by statistical practice. This modification turns out to be enough to relax the restrictions on the rate at which $d$ diverges. The idea behind the mBICV is similar to other modified versions of the BIC that were developed for linear models, see \citep{zak2011modified}.

The mBICV is useful for two things: selecting pair-copulas individually and selecting hyper-parameters of sparse model classes. One such class are \emph{truncated} vine copulas \citep{kurowicka2011, brechmann2012}. A vine copula model is called truncated if all pair-copulas that are conditioned on more than $M$ variables are set to independence. We further propose an alternative class of sparse models, \emph{thresholded vine copulas}. They induce sparsity by setting a pair-copula to independence when the encoded strength of dependence falls short of a threshold. This idea has been around for a long time and heavily used, but the threshold was commonly tied to the $p$-value of an independence test \citep{dissmann2013}. In the more general form, the threshold is a free hyper-parameter. Such classes of sparse models give a computational advantage, since a substantial number of pair-copulas never have to be estimated.

    \section{Background on vine copulas} \label{background}

This section sets up the notation for vine copula models. For a more thorough introduction, see \citep{joe14} and \citep{Stoeber12}.

\subsection{Copulas}

Copulas are models for the dependence between random variables. By the theorem of \citet{sklar1959},  one can express any multivariate distribution $F$ in terms of the marginal distributions $F_1, \dots, F_d$ and a function $C$, called the copula:
\begin{align} \label{eq:sklar}
F(x_1, \dots, x_d) = C\bigl\{F_1(x_1), \dots, F_d(x_d)\bigr\}, \qquad \mbox{for all } \bm x \in \R^d.
\end{align}
If $F$ is the distribution of a random vector $\bm X$, then $C\colon [0, 1]^d \to [0, 1]$ is the distribution of $\bm U = \{F_1(X_1), \dots, F_d(X_d)\}$. This suggests a two-step approach for estimation. Suppose we have observed data $x_{i, j}$, $i = 1, \dots, n$, $j = 1, \dots, d$. Then we can proceed as follows: first, obtain estimates of the marginal distributions, say $\wh F_1, \dots, \wh F_d$; second, set $\wh u_{i, j} =  \wh F_j(x_{i, j})$ and estimate $C$ based on $\wh u_{i, j}$, $i = 1, \dots, n$, $j = 1, \dots, d$. Assuming that derivatives exist, a formula similar to \eqref{eq:sklar} can be derived for the density $f$ of $F$:
\begin{align*}
f(x_1, \dots, x_d) = c\bigl\{F_1(x_1), \dots, F_d(x_d)\bigr\} \times \prod_{k = 1}^d f_k(x_k), \qquad  \mbox{for all }  \bm x \in \R^d,
\end{align*}
where $c$ is the density of $C$ and called the \emph{copula density}, and $f_1,\dots,f_d$ are the marginal densities..

\subsection{Vine copulas}

\begin{figure}
\tikzstyle{VineNode} = [ellipse, fill = white, draw = black, text = black, align = center, minimum height = 1cm, minimum width = 1cm]
\tikzstyle{DummyNode}  = [draw = none, fill = none, text = white] % white node without text for better positioning
\tikzstyle{TreeLabels} = [draw = none, fill = none, text = black] % T_1, T_2, etc.
\newcommand{\labelsize}{\footnotesize} % Size of labels
\newcommand{\yshiftLabel}{-0.3cm}  % Reduce space between labels and edges
\newcommand{\yshiftNodes}{-0.75cm} % Reduce space between lines
\newcommand{\xshiftTree}{0.5cm}    % Space between trees
\newcommand{\rotateLabels}{-57}    % Rotation of labels
	\centering
	\begin{tikzpicture}	[every node/.style = VineNode, node distance =1.4cm] % node distance = scaling of edges
	%%% Tree 1 nodes
	\node (1){1}
	node[DummyNode]  (Dummy12)   [right of = 1]{} 
	node             (2)         [right of = Dummy12] {2}
	node             (3)         [below of = Dummy12, yshift = \yshiftNodes] {3}
	node[DummyNode]  (Dummy45)   [below of = 3, yshift = \yshiftNodes]{}
	node             (4)         [left of = Dummy45] {4}
    node             (5)         [right of = Dummy45] {5}
    %%% Tree 2 nodes
    node             (12)        [right of = 2, xshift = \xshiftTree] {1,2} % 
    node[DummyNode]  (Dummy12x)  [right of = 12]{} 
	node             (13)        [below of = Dummy12x, yshift = \yshiftNodes] {1,3}
	node[DummyNode]  (Dummy45c3) [below of = 13, yshift = \yshiftNodes]{} % Notation c fuer conditioned
	node             (34)        [left  of = Dummy45c3] {3,4}
    node             (35)        [right of = Dummy45c3] {3,5}
    %%% Tree 3 nodes
    node             (15c3)      [right of = 35, xshift = \xshiftTree] {1,5;3}
	node[DummyNode]  (Dummy15c3x)[right of = 15c3]{}
	node             (14c3)      [above of = Dummy15c3x, yshift = -\yshiftNodes] {1,4;3}
	node[DummyNode]  (Dummy23c1x)[above of = 14c3, yshift = -\yshiftNodes]{}
	node             (23c1)      [left of  = Dummy23c1x] {2,3;1}	
	%%% Tree 4 nodes
	node             (24c13)     [right of = Dummy23c1x, xshift = \xshiftTree] {2,4;1,3}
    node             (45c13)     [right of = Dummy15c3x, xshift = \xshiftTree] {4,5;1,3}
    %%% Tree labels
	node[TreeLabels] (T1)        [above of = Dummy12] {$T_1$}
	node[TreeLabels] (T2)        [above of = Dummy12x] {$T_2$}
	node[TreeLabels] (T3)        [above of = 23c1] {\hspace{1.7cm}$T_3$} % T_3 manuell mittiger ausrichten (ausprobieren)		
	node[TreeLabels] (T4)        [above of = 24c13] {$T_4$}	 
	;	    	
	%%% Tree 1 edges   
	\draw (1) to node[draw=none, fill = none, font = \labelsize,
	                    above, yshift = \yshiftLabel] {1,2} (2);
	\draw (1) to node[draw=none, fill = none, font = \labelsize, 
	                    rotate = \rotateLabels, above, yshift = \yshiftLabel] {1,3} (3);   
	\draw (3) to node[draw=none, fill = none, font = \labelsize, 
	                    rotate = \rotateLabels, above, yshift = \yshiftLabel] {3,5} (5);  
	\draw (3) to node[draw=none, fill = none, font = \labelsize, 
	                    rotate = -\rotateLabels, above, yshift = \yshiftLabel] {3,4} (4); 
	%%% Tree 2 edges  
	\draw (12) to node[draw=none, fill = none, font = \labelsize, above, 
	                    rotate = \rotateLabels, above, yshift = \yshiftLabel] {2,3;1} (13);   
	\draw (13) to node[draw=none, fill = none, font = \labelsize, above, 
	                    rotate = \rotateLabels, above, yshift = \yshiftLabel] {1,5;3} (35); 
	\draw (13) to node[draw=none, fill = none, font = \labelsize, above, 
	                    rotate = -\rotateLabels, above, yshift = \yshiftLabel] {1,4;3} (34); 
	%%% Tree 3 edges  
	\draw (23c1) to node[draw=none, fill = none, font = \labelsize, above, 
	                    rotate = \rotateLabels, above, yshift = \yshiftLabel] {2,4;1,3} (14c3);   
	\draw (14c3) to node[draw=none, fill = none, font = \labelsize, above, 
	                    rotate = -\rotateLabels, above, yshift = \yshiftLabel] {4,5;1,3} (15c3); 
	%%% Tree 4 edges 
	\draw (24c13) to node[draw=none, fill = none, font = \labelsize, above, 
	                    rotate = -90, above, yshift = \yshiftLabel] {2,5;1,3,4} (45c13);
	\end{tikzpicture}
	\caption{Example of a regular vine tree sequence.}
	\label{fig:rvine}
\end{figure}
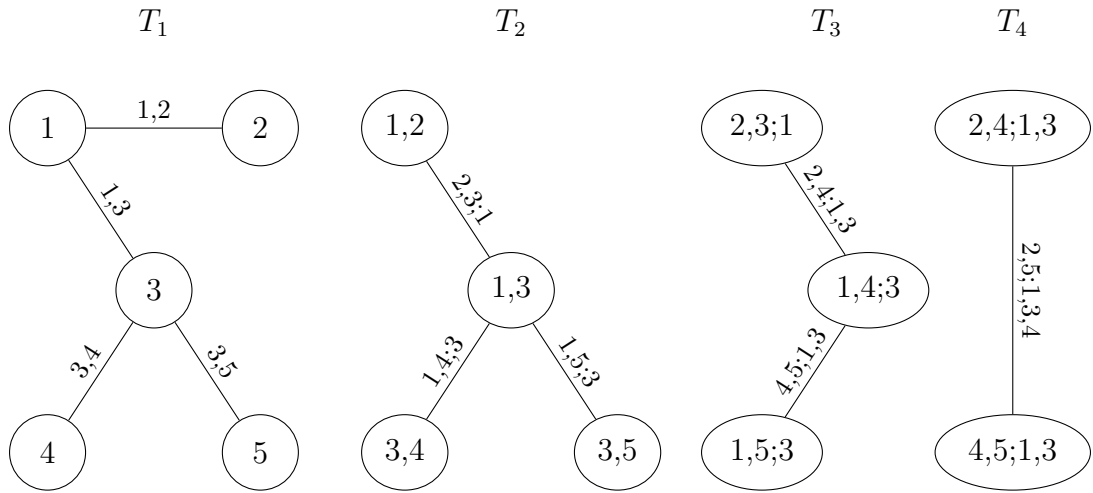

Vine copula models are based on the idea of \citet{joe1996} to decompose the dependence into a cascade of dependencies between (conditional) pairs. However, the decomposition is not unique. \citet{bedford2002} introduced a graphical model, called \emph{regular vine}, that organizes all valid decompositions.  A regular vine on $d$ variables consists of a sequence of linked trees $T_m=(V_m, E_m)$, $m=1, \dots, d-1$ with edge sets $V_m$ and node sets $E_m$ satisfying some conditions, see e.g., \citep{dissmann2013}.
 
A vine copula model identifies each edge in the vine with a bivariate copula. This is best explained by an example. \autoref{fig:rvine} shows a regular vine tree sequence on five variables. The nodes in the first tree represent the random variables $U_1, \dots, U_5$. The edges are identified with a bivariate copula (called \emph{pair-copula}), where edge $(j_e, k_e)$ described the dependence between $U_{j_e}$ and $U_{k_e}$. In the second tree, the nodes are just the edges of the first tree. The edges are annotated by $(j_e, k_e; D_e)$ and describe the  dependence between $U_{j_e}$ and $U_{k_e}$ conditional on $U_{D_e}$. In subsequent trees, the number of conditioning variables increases.

\citet{bedford2001} showed that this naturally leads to a decomposition of the copula density $c$:
\begin{align}
c(\bm u) = \prod_{m=1}^{d-1} \prod_{e \in E_m} c_{j_e, k_e; D_e} \bigl\{G_{j_e|D_e}(u_{j_e}|\bm u_{D_e}), \, G_{k_e|D_e}(u_{k_e}|\bm u_{D_e}) ; \, \bm u_{D_e} \bigr\}, \label{eq:vine_density_ns}
\end{align}
where $\bm u_{D_e}:=(u_\ell)_{\ell \in D_e}$ and $G_{j_e|D_e}$ is the conditional distribution of $U_{j_e} | \bm U_{D_e} = \bm u_{D_e}$. The pair-copula density $c_{j_e, k_e ; D_e}$ is the joint density of the bivariate conditional random vector
\begin{align*}
\bigl(G_{j_e|D_e}(U_{j_e}|\bm U_{D_e}),G_{k_e|D_e}(U_{k_e}|\bm U_{D_e})\bigr)\bigl| \bm U_{D_e} = \bm u_{D_e}.
\end{align*}
\autoref{eq:vine_density_ns} holds for any copula density, but note that the functions $ c_{j_e, k_e; D_e}$ take a third argument $\bm u_{D_e}$. This indicates that, in general, the conditional dependence between $U_{j_e}$ and $U_{k_e}$ may be different for different values of $\bm u_{D_e}$. To facilitate tractability of the models, it is customary to ignore this influence and simplify the model to
\begin{align*}
c(\bm u) = \prod_{m=1}^{d-1} \prod_{e \in E_m} c_{j_e, k_e; D_e} \bigl\{G_{j_e|D_e}(u_{j_e}|\bm u_{D_e}), \, G_{k_e|D_e}(u_{k_e}|\bm u_{D_e}) \bigr\}. %\label{eq:vine_density}.
\end{align*}
In this case the pair copula density  $c_{j_e, k_e; D_e} \bigl\{G_{j_e|D_e}(u_{j_e}|\bm u_{D_e}), \, G_{k_e|D_e}(u_{k_e}|\bm u_{D_e}) \bigr\}$ encodes the \emph{partial dependence} between  $U_{j_e}$ and $U_{k_e}$ conditional on $\bm U_{D_e}$ \citep{gijbels2015, spanhel2015a}. This \emph{simplifying assumption} is often valid for financial data and even if it is violated, it may serve as a useful approximation to the truth. For further discussions, see \citep{spanhel2015b, nagler2016, acar2012,stoeber2013,kraus2017growing,kurz2017testing,vatter2018generalized}, and references therein. Since tractability is vital in very high dimensions, we shall assume that the simplifying assumption holds for the remainder of this paper.

    \section{A modified BIC for sparse vine copula models} \label{sec:mBIC}

%The thresholded vine copula model introduced in the previous sections is characterized by a hyper-parameter $\theta$ that controls the degree of sparsity. Depending on the choice of this hyper-parameter, the models can range between a full model with no independence and the independence model. The natural question is: what is the right amount of sparsity? In some cases, expert knowledge can be used to determine an appropriate value for the hyper-parameter. But usually, it is desirable to make this choice automatic and data-driven. 
%
%In the following, we propose a procedure that is specifically tailored to high dimensional applications. It is based on a  \emph{generalized information criterion(GIC)} that, unlike the AIC or BIC, assumes that the number of possible parameters is larger than the sample size. 

Consider a vine copula model with density $c^{\bm \eta}$ that is characterized by a finite dimensional parameter $\bm \eta$. We denote the number of parameters in the model by $\nu$. To simplify our arguments, we assume that there is only one parameter per pair-copula, i.e., $\bm \eta = (\etae)_{e \in E_m, m = 1, \dots, d-1}$, and that $\eta_e$ is zero if and only if the pair-copula at edge $e$ is independence. However, we formally distinguish between the number of non-zero parameters $\nu$ and the number of non-independence pair-copulas $q$ to maintain validity of the formulas in the general context.  

The maximal model contains $\kmax = d(d-1)/2$ non-independence copulas.  A sparse vine copula model contains many independence copulas and, thus, only few non-zero parameters. Let $q = \#\{\eta_e\colon \eta_e \neq 0, e \in E_{1}, \dots, E_{d - 1}\}$ be the number of non-independence copulas in the model. We speak of a sparse model if $q \ll \kmax$ or, asymptotically, $q / \kmax \to 0$.

\subsection{The BIC and why it is inappropriate for sparse models}
\label{sec:bic}

Let $\bm u_i \in [0, 1]^d$, $i =1, \dots, n$, be \emph{iid} observations and $\wh{\bm \eta}$ be the estimated parameter. The \emph{Bayesian information criterion (BIC)} is defined as
\begin{align*}
\mathrm{BIC}(\etah ) =  -2  \ell({\etah}) + \wh \nu \ln n,
\end{align*}
where $\ell({\etah}) = \sum_{i=1}^n  \ln c^{\etah}\bigl(\bm u_i\bigr)$ denotes the log likelihood and $\wh \nu$ is the number of non-zero parameters in $\etah$. The lower the BIC, the more favorable we see a model. 

The BIC of a vine copula model decomposes to
\begin{align*}
&\phantom{=}\; \mathrm{BIC}(\etah ) \\
&= -2  \sum_{m=1}^{d-1} \sum_{e \in E_m} \sum_{i = 1}^n \bigl[ \ln c_{j_e, k_e; D_e}^\etahe \bigl\{G_{j_e|D_e}^{\etah_{D_e}}(u_{i, j_e}|\bm u_{i, D_e}), \, G_{k_e|D_e}^{\etah_{D_e}}(u_{i, k_e}|\bm u_{i, D_e}) \bigr\} + \wh \nu_e \ln n \bigr]  \\
&= -2  \sum_{m=1}^{d-1} \sum_{e \in E_m} [\ell_e(\etahe, \etah_{D_e}) + \wh \nu_e \ln n \bigr]
\end{align*}
where $\wh \nu_e$ is the number of parameters for edge $e$ and 
\begin{align*}
\ell_e(\etahe, \etah_{D_e}) = \sum_{i = 1}^n \bigl[ \ln c_{j_e, k_e; D_e}^\etahe \bigl\{G_{j_e|D_e}^{\etah_{D_e}}(u_{i, j_e}|\bm u_{i, D_e}), \, G_{k_e|D_e}^{\etah_{D_e}}(u_{i, k_e}|\bm u_{i, D_e}) \bigr\}. 
\end{align*}
This decomposition suggests that the global BIC can be minimized by sequentially minimizing the BIC of individual pair-copulas.  But there is a subtle issue: 
the log likelihood $\ell_e$ for an edge in tree $e \in E_{m}$, $m \ge 2,$ also depends on estimated parameters $\etah_{D_e}$ from previous trees. 
In general, step-wise and global optima may correspond to different models. But both criteria select the same model as $n \to \infty$. This motivates the common practice of selecting the pair-copulas individually in sequential estimation.

The BIC is derived from a Bayesian argument that states that the posterior log probability of a model is proportional to
\begin{align} \label{eq:BIC}
- 2 \ell(\etah) + \wh \nu \ln n - 2 \ln \psi({\etah}) + O_p(n),
\end{align}
where $\psi(\etah)$ is the prior probability of the model $c^\etah$ and the $O_p(n)$ term is independent of the model choice.
The simpler form of the BIC is then a consequence of the assumption that all models are equally likely \emph{a priori}. When $\kmax$ (hence $d$) is fixed, it is well known that the BIC selects the true model with probability going to 1 as $n$ tends to infinity; see, e.g., \citep{claeskens2008model}. In high-dimensional vine copula models, $\kmax$ is often of the same order or much larger than $n$. In such situations, ``large $n$, fixed $d$'' asymptotics are at least questionable. 

Since all models are equally likely, we expect the number of parameters in the model to be $\kmax / 2$ which stands in contrast to the sparsity assumption $q / \kmax \to 0$. In fact, one can show with arguments similar to those in \autoref{sec:consistency} that the BIC cannot distinguish the true from an incorrect model when $\kmax \ge \sqrt{n \ln n}$ or equivalently $d \ge \sqrt[4]{n \ln n}$. In terms of the number of variables $d$, this restriction is much more severe for vine copula models compared to (generalized) linear models, where the BIC has been studied most extensively \citep{zak2011modified}.

\subsection{A modified criterion} \label{sec:mBICV}

Several authors considered modifications of the BIC to make it more suitable for high-dimensional problems \citep{Bogdan989, chen2008, zak2011modified}. The unifying idea is to adjust the prior probabilities in \eqref{eq:BIC} such that sparse models are more likely than dense models. We shall follow the same path and assign each pair-copula a prior probability $\psi_e$ of not being independent. 

More precisely, we assume that the indicators $I_e = \ind(\eta_e \neq 0)$ are independent Bernoulli variables with mean $\psi_e$ and propose to choose $\psi_e = \psi_0^m$ for any edge in tree $m$. The resulting prior probability of a vine copula model $c^{\bm \eta}$ is
\begin{align*}
\psi(\bm \eta) = \prod_{m = 1}^{d - 1} \psi_0^{m q_m}(1 - \psi_0^m)^{d - m - q_m},
\end{align*}
where $q_m$ is the number of non-independence copulas in tree $m$.
Now \eqref{eq:BIC} suggests the criterion
\begin{align*}
\mathrm{mBICV}({\etah}) 
= - 2 \ell (\etah) + \wh \nu \ln n - 2\sum_{m = 1}^{d - 1} \bigl\{\wh q_m \ln \psi_0^m + (d - m - \wh q_m)\ln(1 - \psi_0^m)\bigr\},
\end{align*}
where $\ell (\etah) = \sum_{i = 1}^n c^\etah(\bm u_i)$
is the log-likelihood, and $\wh q_m$ is the number of non-independence copulas in tree $m$ of the fitted model $\etah$.
The mBICV decomposes similarly to the BIC:
\begin{align}
&\phantom{=}\; \mathrm{mBICV}({\etah} ) \notag \\
&=  \sum_{m=1}^{d-1} \sum_{e \in E_m} \biggl[  -2 \ell_e(\etahe, \etah_{D_e}) + \wh \nu_e \ln n  - 2\bigl\{\widehat  I_e \ln \psi_0^m + (1 - \widehat I_e)\ln(1 - \psi_0^m)\bigr\} \biggr] \notag \\
&=  \sum_{m=1}^{d-1} \sum_{e \in E_m} \mathrm{mBICV}_e\bigl(\etahe, \etah_{D_e}\bigr), 
\label{eq:mBICV_decomposition}
\end{align}
suggesting that $\mathrm{mBICV}$ is a suitable criterion for sequential selection of the pair-copulas.

A decreasing sequence $\psi_0^m$ implies that higher-order pairs are more likely to be (conditionally) independent. Hence, mBICV penalizes non-independence copulas more severely compared to the BIC, but only in higher trees. In the first tree, the prior probability of a non-independence copulas is $\psi_0$. If $\psi_e > 0.5$, the mBICV is more likely to select a non-independence model. For $\psi_0 = 0.9$, this means that the first seven trees are more favorable to non-independence compared to BIC. This feature is motivated by statistical practice. All popular structure selection algorithms try to capture strong dependence relationships in the first few trees \citep{dissmann2013, muller2017selection, brechmann2015truncation, kraus2017growing}.  Although there is no guarantee, this typically leads to models where there is a lot of dependence in the first few trees and only little dependence in higher trees. Further, the parameters of a vine copula are typically estimated sequentially starting from the first tree. Because estimation errors accumulate over the trees, estimates in higher trees are less reliable. And the less reliable the estimates are, the more conservatively we should choose our model.

Under the prior used for the mBICV, $q_m$ is a binomial experiment of $(d - m)$ trials with success probability $\psi_0^m$. Therefore, the expected number of non-independence copulas in tree $m$ is $E(q_m) = (d - m)\psi_0^m$ and the expected total number is
\begin{align} \label{eq:edgesum}
E(q) = \sum_{m = 1}^{d- 1} (d - m)\psi_0^m = O\biggl(d\sum_{m = 1}^{d- 1}\psi_0^m\biggr) = O(d).
\end{align}
Recalling that $\kmax \sim d^2$, we obtain  $E(q) / \kmax \to 0$ and, hence, expect the model to be sparse. 

A referee noted that there are other choices for $\psi_e$ that correspond to sparse models. For example, $\psi_e = \psi_0/m$ leads to $E(q) = O(d \ln d)$ which induces sparsity at a slightly slower rate. In that sense, our choice $\psi_e = \psi_0^m$ is somewhat arbitrary. In the following theoretical arguments the two choices are essentially equivalent. But since sparser models are also computationally more tractable, inducing sparsity at a faster right seems reasonable.

    \section{Asymptotic properties} \label{sec:consistency}

We will argue that the mBICV can consistently distinguish between a finite number of models provided that $d = o(\sqrt{n})$, which is a less stringent condition compared to BIC. This is only a weak form of consistency: when $d \to \infty$, it does not automatically imply that the criterion finds the best among \emph{all possible} candidate models. Because the number of possible candidates grows rapidly with $d$, results on this stronger form of consistency are considerably more difficult to derive and have only recently emerged in the simpler context of linear models \citep{wang2011, wang2009, fan2013}.

Unfortunately, asymptotic results for parameter estimates in vine copula models of diverging size are not yet available. For the Gaussian copula, \citet{xue2014} showed that the parameters can be estimated consistently when $d$ diverges, although at a rate slower than $O_p(n^{-1/2})$. The Gaussian copula is a special case of a vine copula, where all pair-copulas are Gaussian and $\eta_e$ equals the partial correlation $\rho_{j_e, k_e;D_e}$. Some general results for (joint) maximum-likelihood estimation were derived in \citep{fan2004}, but do not cover the sequential estimation method that is typically used for vine copula models. 

This lack of results makes it impossible to formally establish consistency of the mBICV. Instead, we give a semi-formal argument that explains why we expect the mBICV to be consistent when $d = o(\sqrt{n})$ and substantiate this claim by a small simulation experiment. 
We shall consider the decomposition \eqref{eq:mBICV_decomposition} and approximate the probabilities of selecting a wrong model for a fixed edge in tree $m < \infty$. The resulting probabilities of selecting the wrong model for a given edge are then aggregated and shown to vanish asymptotically. Since we do not approximate the error probabilities uniformly in $m$ and neglect lower-order terms when applying the central limit theorem, the argument is only heuristic and should be taken with a grain of salt.

\subsection{An informal argument}

Our null hypothesis is that the true pair-copula at edge $e \in E_m$ corresponds to independence, i.e., $\eta_e = 0$. Denote by $\alpha_{n, e}$ the probability of a type I error (selecting the non-independence model $\etae \neq 0$ although the true model has $\etae = 0$), i.e.,
\begin{align*}
\alpha_{n, e} = \Pr\bigl\{\mathrm{mBICV}_e\bigl(\etahe, \etah_{D_e}\bigr) < \mathrm{mBICV}_e\bigl(0, \etah_{D_e}\bigr) \mid \eta_e = 0\bigr\},
\end{align*}
and by $\beta_{n, e}$ the probability of a type II error (selecting the independence model $\eta_e = 0$ although the true model has $\eta_e \neq 0$), i.e.,
\begin{align*}
\beta_{n, e} &=  \Pr\bigl\{\mathrm{mBICV}_e\bigl(\etahe, \etah_{D_e}\bigr)  > \mathrm{mBICV}_e\bigl(0, \etah_{D_e}\bigr) \mid \etae \neq 0 \bigr\}.
\end{align*}
In \autoref{sec:proof1} we argue that for any $e \in E_m$, $m < \infty$, the error probabilities are approximately
\begin{align} \label{eq:error_probabilites}
 \alpha_{n, e} = O\biggl( \frac{\psi_0^m}{\sqrt{n \ln n}} \biggr), \qquad
 \beta_{n, e} = O\biggl(\frac{e^{-nMI_e^2}}{\sqrt{n}MI_e}\biggr), 
\end{align}
where 
\begin{align*}
\mathrm{MI}_e &= \int_{[0, 1]^2} c_{j_e, k_e; D_e}^{\eta_e}(u, v) \ln c_{j_e, k_e; D_e}^{\eta_e}(u, v) du dv,
\end{align*}
is the mutual information associated with edge $e$. 

The type I error probability  $\alpha_{n, e}$ is increasing in $\psi_0^m$, the prior probability of a non-independence model in tree $m$. Lower values shift our expectations to sparser models and the probability of overfitting decreases. In particular, since $\psi_0^m$ is a decreasing sequence, $\alpha_{n, e}$ is decreasing in the tree level $m$. This is related to the fact that the mBICV expects more independence copulas at higher tree levels. The type II error probability $\beta_{n, e}$ decreases in the mutual information $\mathrm{MI}_e$ which confirms our intuition that stronger dependence relationships are easier to detect. The rate for $\beta_{n, e}$ is unlikely to hold uniformly in $m$, since our argument assumes $\| \etah_{D_e} - \etab_{D_e} \|_2 = O_p(n^{-1/2})$, which is only valid when $m$ is fixed. However, $\beta_{n, e}$ vanishes much faster than $\alpha_{n, e}$ and we conjecture that this is still the case when taking $m \to \infty$. The qualitative interpretation remains valid in any case. 

The error probabilities in \eqref{eq:error_probabilites} can now be aggregated over the whole vine.
Consider the family-wise error rates (FWER), i.e., the probability of selecting the wrong model for at least one edge in the model. Define  $\alpha_{n}$ as the FWER of a type I and similarly $\beta_n$ for the type II error. Using Bonferroni's inequality, we get 
\begin{align*}
\alpha_{n} \le \sum_{m = 1}^{d - 1} \sum_{e \in E_m\colon \eta_e = 0}\alpha_{n, e}, \qquad  
\beta_{n} \le \sum_{m = 1}^{d - 1} \sum_{e \in E_m\colon \eta_e \neq 0}\beta_{n, e}.
\end{align*}
By expanding the sum and using \eqref{eq:error_probabilites}, we obtain 
\begin{align*}
\alpha_n \le \frac{1}{\sqrt{n \ln n}}\sum_{m = 1}^{d - 1} (d - m)\psi_0^m = O\biggl(\frac{d}{\sqrt{n \ln n}} \biggr).
\end{align*}
Similarly, we get  $\beta_n = O(d^2e^{-nMI_e^2} / \sqrt{nMI_e}) = o(\alpha_n)$. Combining the results for type I and type II suggest that the mBICV is consistent when $d = o\bigl(\sqrt{n \ln n}\bigr)$. However, our approximation for the type I error requires $\| \etah - \etab \|_2 = o_p(1)$. Each non-zero parameter can be estimated at best at rate $O_p(n^{-1/2})$ which suggests that $\| \etah - \etab\|_2 = O_p(d / \sqrt{n})$. Hence, we must strengthen our condition for consistency of the mBICV to $d = o(\sqrt{n})$.

\subsection{Simulation}

\begin{figure}[t]
\centering
\includegraphics[width = 0.9\textwidth]{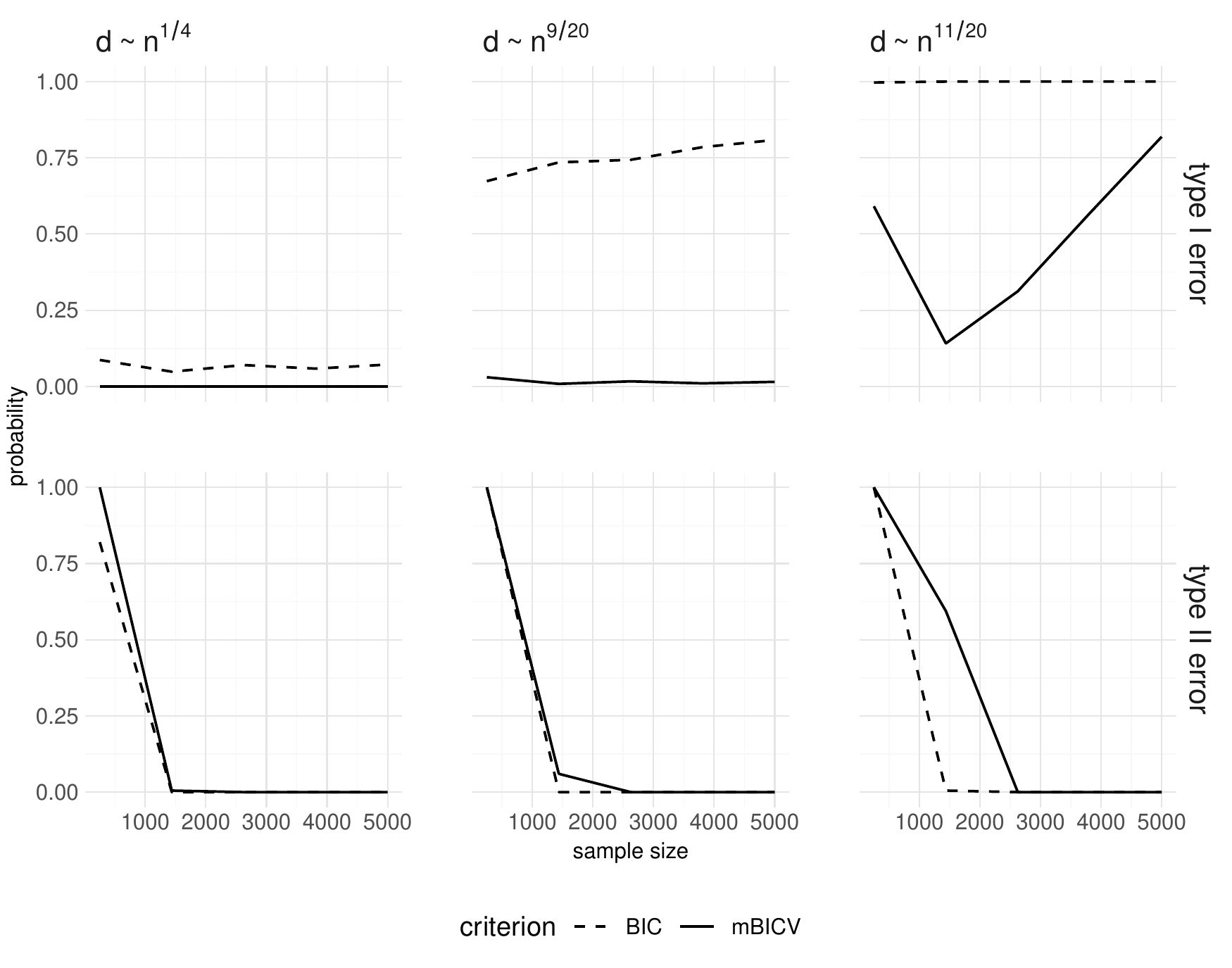}
\caption{Family-wise error rates with simultaneously increasing sample size and dimension in three asymptotic regimes.} \label{fig:probs}
\end{figure}

Since the arguments above are only heuristic, we conduct a small simulation experiment to check whether the condition $d = o(\sqrt{n})$ is reasonable. We consider a simple setup where the true model contains only Gaussian pair-copulas. Half of the pair-copulas in every tree are set to independence ($\eta_e = 0$), all others have parameter $\eta_e = 0.2$. Note that this model corresponds to the implicit prior probabilities by the BIC and is not sparse. We simulate data sets of increasing sample size $n$ and let the dimension $d$ grow with $n$ at three different rates. We consider five sample sizes and simulate 1\,000 data sets each. We always fit two models: one selects the copula families based on BIC, the other based on mBICV. 

\autoref{fig:probs} shows the probabilities of selecting at least one pair-copula incorrectly, differentiating between type I ($\alpha_n$) and type II ($\beta_n$) errors. The three columns correspond to the three asymptotic regimes $d \sim n^{1/4}$, $d \sim n^{9/20}$, and $d \sim n^{11/20}$, respectively.
First, we observe that the type II error probabilities  vanish fast for both criteria. This is in line with our asymptotic arguments that suggested that type II errors are much easier to avoid. The type I error probability is the main factor determining consistency of the criteria.

In the first regime ($d \sim n^{1/4}$), we expect both BIC and mBICV to consistently select the true model as $n$ becomes large. The type I error probability of the mBICV is zero for every sample size under consideration. The error probability of the BIC decreases slowly, which is also in line with our theoretical considerations: the BIC is consistent when $d = o (\sqrt[4]{n \ln n})$ and $d \sim n^{1/4}$ is close to the border of that range. In the second regime, we still expect the mBICV to be consistent since $d \sim n^{9/20} = o(\sqrt{n})$, but not the BIC. This is supported by the simulations: while the type I error probability vanishes for mBICV, it is increasing for the BIC. Finally, we expect both criteria to be inconsistent for $d \sim n^{11/20}$ and, indeed, the type I error probabilities tend to one for both criteria. In summary, the simulations support our hypothesis that the mBICV is consistent when $d = o(\sqrt{n})$.

    \section{Special classes of sparse vine copula models and their selection} \label{sec:sparse-vines}

The mBICV can be used to decide which pair-copulas in the model are set to independence. But in order to calculate the mBICV, one first needs to estimate a model. In high-dimensional vine copula models, there is a huge number of pair-copulas. Estimating all of them will be computationally demanding. This predicament is solved by focusing on sparse model classes that set a large proportion of pair-copulas to independence before a model is fit. Below we discuss two such classes, \emph{truncated} and \emph{thresholded} vine copula models. Both classes have a sparsity inducing hyper-parameter that can be selected by mBICV.

\subsection{Truncated vine copulas}

Truncated vine copula models induce sparsity by setting all pair-copulas after a certain tree level $M \in \{1, \dots, d-1\}$ to the independence copula. The lower the \emph{truncation level} $M$, the higher the degree of  sparsity: an $M$-truncated vine copula model allows for only $M(2d - M - 1)/2$ non-independence copulas. Since the density of the independence copula is $1$ everywhere, the density of an $M$-truncated vine copula model can be written as
\begin{align*}
c(\bm u) = \prod_{m=1}^{M} \prod_{e \in E_m} c_{j_e, k_e; D_e} \bigl\{G_{j_e|D_e}(u_{j_e}|\bm u_{D_e}), \, G_{k_e|D_e}(u_{k_e}|\bm u_{D_e}) \bigr\}. 
\end{align*}

The logic behind truncated vine copulas is closely related to the structure selection heuristic of \citet{dissmann2013}. Its goal is to capture most of the dependence in the first couple of trees. If this would allow to capture \emph{all} dependence in the first $M$ trees, the truncated model arises naturally. If this is not the case, one can at least hope that the dependence in higher trees is practically irrelevant. 

\subsection{Thresholded vine copulas} \label{sec:thresh-models}

The idea of thresholded regular vines is different. The ultimate goal is to set all conditional pair-copulas that are practically irrelevant to independence. Arguably, `practical relevance' is a vague concept and we need to rely on a proxy measure for it. A natural choice are measures for the strength of dependence. On such measure is  Kendall's $\tau$. It is a \emph{measure of concordance} and, as such, is a functional of the copula only. In particular, it does not depend on the marginal distributions \citep{nelsen2006}. In the remainder of this article we take Kendall's $\tau$ as our target measure, although other choices are equally valid.

Similar to the truncation level $M$ before, thresholded vine copulas have a hyper-parameter $\theta$, called \emph{threshold}. Denote the Kendall's $\tau$ associated with the pair-copula $c_{j_e, k_e; D_e}$ by $\tau_e$ and define $E_m^\theta = \{e \in E_m: \vert \tau_e \vert > \theta \}$ for any edge set $E_m$ in the vine. Then the the density of a thresholded vine copula model becomes
\begin{align*}
c(\bm u) = \prod_{m=1}^{d - 1} \prod_{e \in E_m^\theta} c_{j_e, k_e; D_e} \bigl\{G_{j_e|D_e}(u_{j_e}|\bm u_{D_e}), \, G_{k_e|D_e}(u_{k_e}|\bm u_{D_e}) \bigr\}. 
\end{align*}
While it would be possible to use a different threshold in each tree, we shall only consider the simple case where $\theta$ is constant across all tree levels.

The number of non-independence copulas in the model can be controlled by the threshold parameter $\theta$. But in contrast to the truncated model, the number also depends on the actual dependence in the random vector $\bm U$. To illustrate this, we fix $\theta$ and consider a few interesting boundary cases:
\begin{itemize}
\item  If $\vert \tau_e \vert > \theta$ for all $e \in E_1, \dots, E_{d-1}$,  the thresholded model is equal to the full model. 
\item If for $1 \le M < d-1$, it holds $\vert \tau_e \vert > \theta$ for all $e \in E_1, \dots, E_{M}$ and $\vert \tau_e \vert \le \theta$ for all $e \in E_M, \dots, E_{d_1}$, the thresholded model is equal to the $M$-truncated model.
\item  If $\vert \tau_e \vert \le \theta$ for all $e \in E_1, \dots, E_{d-1}$,  the thresholded model contains only independence pair-copulas and, thus, is equal to the independence model.
\end{itemize}
The thresholded model generalizes the truncated model: whenever a model is truncated at level $M$, it can be represented as a thresholded model with $\theta < \min_{e \in E_1, \dots E_{M}\colon \tau_e \neq 0} \vert \tau_e\vert$. If a model is $M$-truncated, it holds that $\tau_e = 0$ for all $e \in E_m, m > M$. If we choose, e.g., $\theta = \min_{e \in E_1, \dots E_{M}\colon \tau_e \neq 0} \vert \tau_e\vert / 2$, then $\theta < \vert \tau_e \vert$ for all non-independence copulas up to tree $M$ and $\theta > \vert \tau_e \vert$ for all remaining pairs. Hence, the thresholding operation keeps all non-independence copulas of the truncated model and sets all others to independence. If the original model is not truncated, the thresholded model can still adapt to the sparsity patterns of the truth and is therefore more flexible.

The thresholding idea is not new. Several authors used a similar idea, but tied the value of $\theta$ to the critical value of a significance test for an empirical version of Kendall's $\tau$ \citep{czado2012, brechmann2012, dissmann2013}. The idea is to perform a statistical test for independence and set the pair-copula to independence whenever the null hypothesis cannot be rejected. This procedure is only heuristic: in general, the test based on the empirical Kendall's $\tau$ is not consistent for the null hypothesis of independence and there is no correction for multiple testing. Treating the threshold as a free hyper-parameter brings additional flexibility and allows to tailor the threshold to a specific application.

\subsection{Automatic hyper-parameter selection based on mBICV}
\label{sec:auto}

The mBICV allows us to compare models for various thresholds and decide which is the best. The natural way to select the best model is to fit several models for a fine grid of $\theta$ values and select the one with lowest mBICV. This strategy can be extremely time-consuming in high dimensions, where a single fit of the full model often takes hours. An advantage of sparse vine copula models is that only a fraction of the pair-copulas needs to be estimated (all others are set to independence). This motivates the following strategy to automatically select the threshold parameter $\theta$: 
\begin{enumerate}[1.]
\item Start with a large threshold and fit the model. 
\item Reduce the threshold and fit a new model.
\item If the new model improves the mBICV, continue with 2. Stop if there is no improvement.
\end{enumerate}

This approach has several computational advantages. Since we start with a large threshold, only a few pair-copulas have to be estimated in the first iteration. For all following iterations, only a few pair-copulas will change from one iteration to the next. By keeping the result from the previous fit in memory, we can re-use most of the fitted pair-copulas and only need to estimate a few additional parameters. Upon termination of the algorithm, most of the non-independent pair-copulas have only been estimated once. Thus, the overall time spent on model fitting and selection will be comparable to the time required for fitting \emph{only} the mBICV-optimal model. Depending on the dimension $d$ and the level of sparsity in the data, this can be several times faster than fitting the full model. 

The same strategy can be used for selecting the truncation level: start with a low truncation level and gradually add more tree levels until the mBICV fails to improve. It is similarly straightforward to select the threshold and the truncation level simultaneously by using an outer loop for the threshold and an inner loop for the truncation level.

\subsection{Implementation}

We propose to reduce the threshold in a data-driven manner. For the initial threshold, we choose the maximum of all pair-wise absolute empirical Kendall's $\tau$. In this case, the initial model consists of only independence copulas and $\mathrm{mBICV} = 0$. To reduce the threshold from one iteration to the next, we choose the threshold such that about $5\%$ of the previously independent pairs may become non-independent in the new model. To be more precise, let $\theta_k$ denote the threshold in iteration $k$, $\mathcal{T}_k = \{\tau_e: \vert \tau_e \vert \le \theta_k\}$ and $N_k = \vert \mathcal{T}_k \vert$. Then we set $\theta_{k + 1}$ to the $\lceil 0.05 N_k\rceil$-th largest value in $\mathcal{T}_k$. 

To check if a pair-copula from the last iteration can be re-used, we need to check if the pseudo-observations have changed. In very high-dimensional models it is also important to make efficient use of memory. We should not store the pseudo-observations of each conditional pair in the vine. Because of the large number of pair-copulas, this will quickly exceed the memory of customary computers. Instead, one can store a summary statistic (like a weighted sum) of the pseudo-observations and only check if this summary statistic has changed.

An open-source implementation of the selection algorithm in \texttt{C++} is available as part of the \texttt{vinecopulib} library  and its  interface to \textsf{R} \citep{rvinecopulib}.

    \section{Case study: Modeling dependence in a large stock portfolio}

To illustrate the concepts introduced in the previous section, we use the new methodology to model the dependence between a large number of stocks. Our data set contains daily stock returns from the S\&P 100 constituents from the period of 1 January 2010 to 31 December 2016 ($n = 1756$). Only stocks that were traded over the whole period are included, leaving us with $d = 96$. 

\subsection{Modeling} \label{sec:appl:method}

\subsubsection*{Models for the marginal time series}

A stylized fact about stock returns is that the squared time-series exhibit strong inter-serial dependence. A popular model that takes this fact into account is the ARMA-GARCH model \citep{francq2011garch}.
Let $x_{t, k}$, $t = 1, \dots, T$ be the returns of stock $k$ at time $t$. The ARMA(1, 1)-GARCH(1, 1) model for this time series is
\begin{align*}
x_{t, k} &= \mu_k  + \phi_k x_{t - 1, k} + \psi_k a_{t-1, k}  + a_{t, k} \\ 
a_{t, k} &= \sigma_{t, k}\epsilon_{t, k} \\
\sigma_{t, k}^2 &= \omega_k  + \beta_k \sigma_{t - 1, k}^2 + \alpha_k \epsilon_{t - 1, k}^2,
\end{align*}
where $\epsilon_{t, k}$ are \emph{iid} Student t variables with zero mean, unit variance, and $\nu_k$ degrees of freedom. The parameters of this model can be estimated with maximum likelihood, for example using the R package \texttt{fGarch} \citep{fGarch}. In addition to parameter estimate, this also gives us estimates $\wh \epsilon_{t, k}$ and $\wh \sigma_{t, k}$ of the unobserved innovation process $\epsilon_{t, k}$ and volatility process $\sigma_{t, k}$. 

\subsubsection*{Dependence model}

The cross-sectional dependence between stocks is modeled by a vine copula underlying the residual time series $\bm \epsilon_t = (\epsilon_{t, 1}, \dots, \epsilon_{t, d})$, $t = 1, \dots, T$. We assume that the dependence in $\bm \epsilon_t$ is induced by a vine copula model $C$. Define $u_{t, k} = \Psi(\epsilon_{t, k}; \nu_k)$, where $\Psi(\cdot; \nu_k)$ is the cumulative distribution function of a Student t distribution with zero mean, unit variance, and degrees of freedom $\nu_k$. Then, $\bm u_t = (u_{t, 1}, \dots, u_{t, d})$, $t = 1, \dots, T$, are independent observations from a random vector with distribution $C$. 

In practice, we do not observe $\bm u_t$. Suppose that $\wh{\bm \epsilon}_t$ are the observed residual series and $\wh \nu_k$ are the estimated degrees of freedom parameters of the fitted ARMA-GARCH models. Then $\wh u_{t, k} =  \Psi(\wh \epsilon_{t, k}; \wh \nu_k)$, $k = 1, \dots, d$, $t  = 1, \dots, T$, act as pseudo-observations of the copula $C$. Based on these, a thresholded vine copula model can be estimated for any fixed value of the threshold $\theta$. We only allow for parametric pair-copulas, and choose the family of each pair-copula from the Gaussian, Student $t$, Clayton, Gumbel, Frank, and Joe families by the mBICV criterion. The prior probability $\psi_0$ is set to 0.9 and the vine structure is selected by the algorithm of \citet{dissmann2013}.

\subsection{Illustration of the new concepts}

\begin{figure}[h!]
\centering
\subfloat[selection based on mBICV]{
\includegraphics[width = 0.49\textwidth]{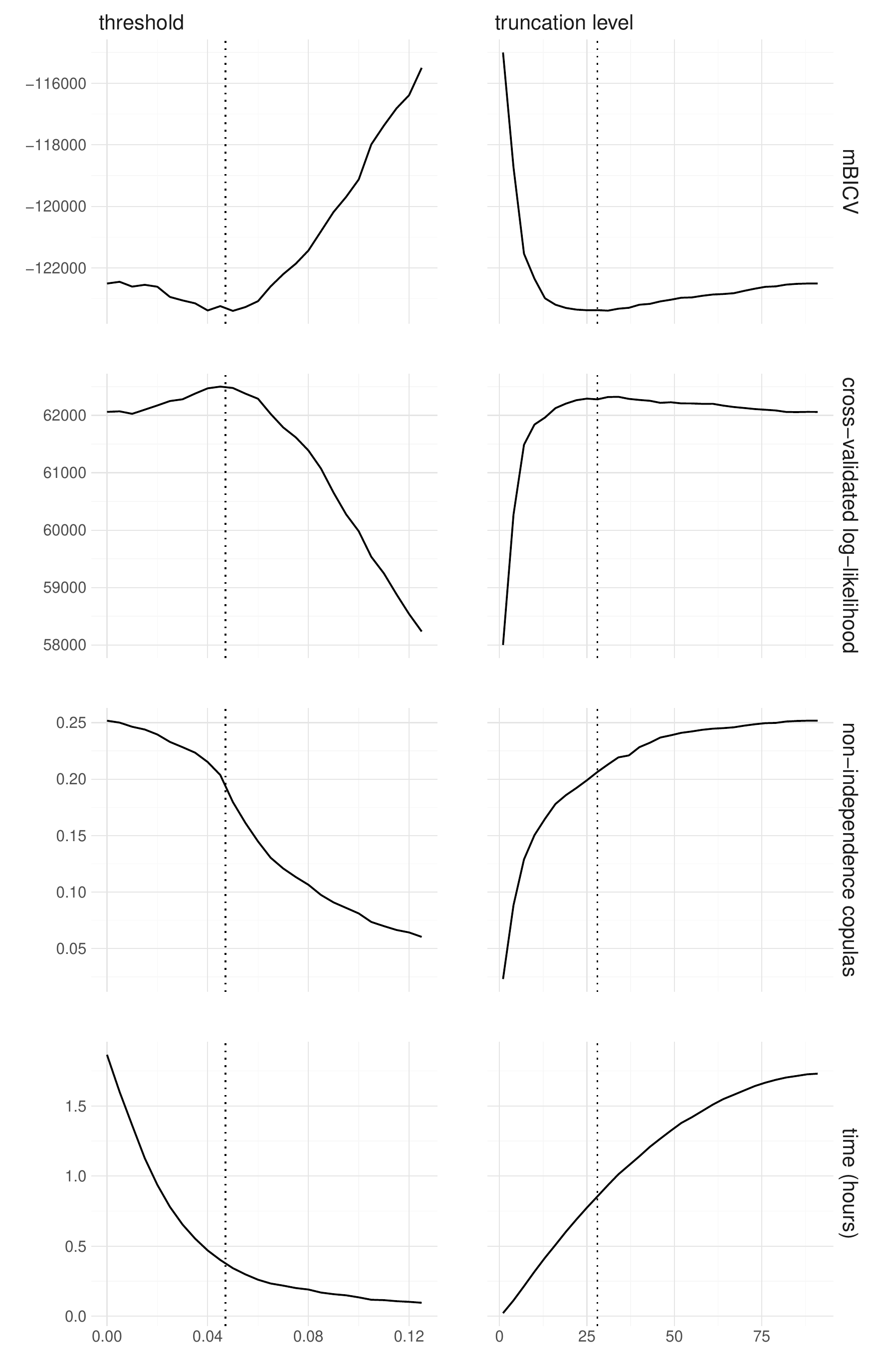} \label{fig:mbicv}
}
\subfloat[selection based on BIC]{
\includegraphics[width = 0.49\textwidth]{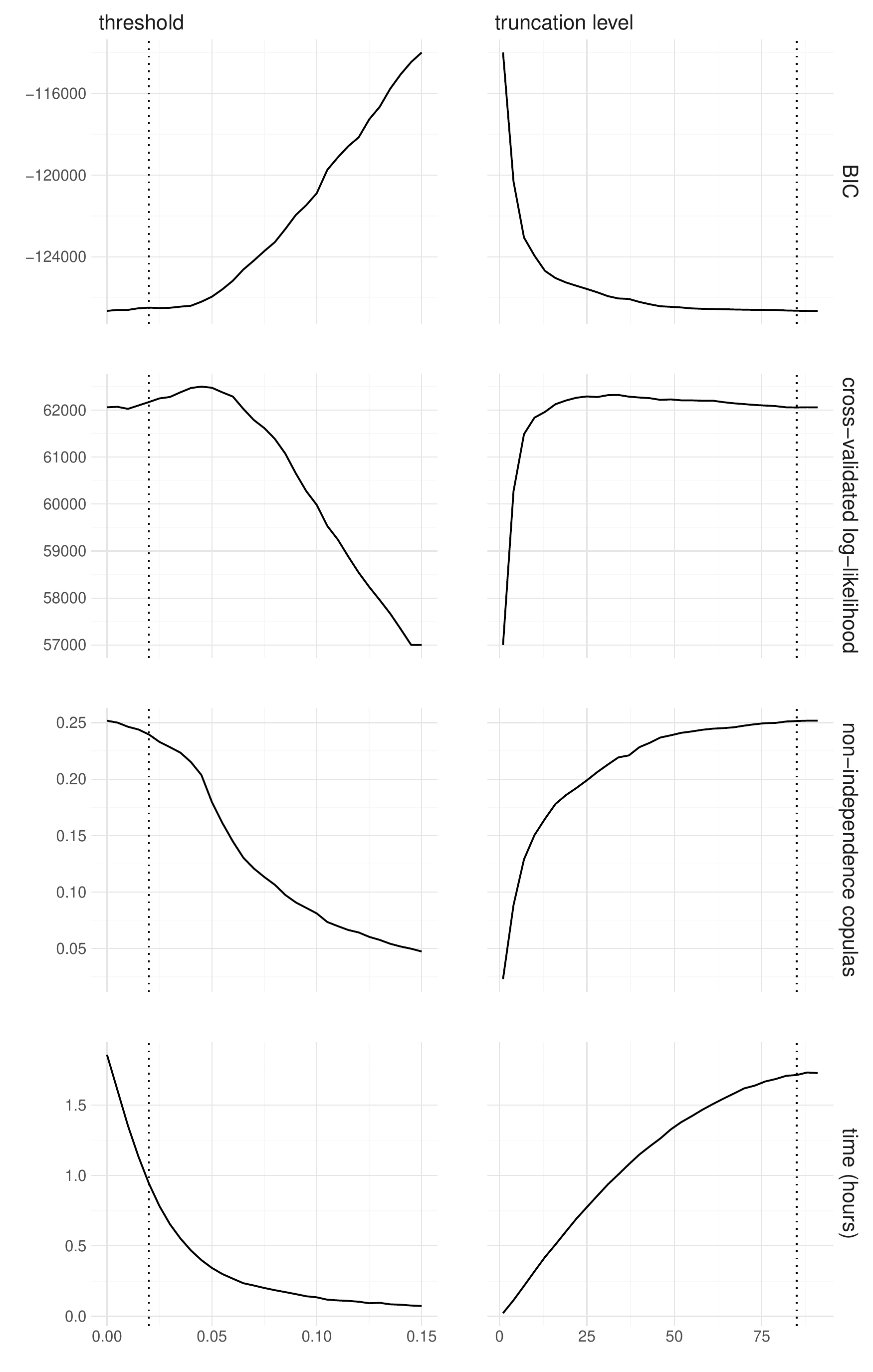} \label{fig:bic}
}
\caption{The selection criterion, cross-validated log-likelihood, proportion of non-independence copulas, and time required to fit the model as functions of the sparsity hyper-parameters. The dotted lines indicate the values of the hyper-parameter that have been selected by the automatic selection procedure from \autoref{sec:auto}.}
\end{figure}

To assess the influence of the sparsity hyper-parameters in thresholded and truncated vine copula models, we shall first estimate the marginal and dependence models using the full data from 1 January 2010 to 31 December 2016. This allows us to make a direct connection between the hyper-parameters and the mBICV, which helps us to understand their relation. In particular, we fit vine copula models to the pseudo-observations $\wh{\bm u_t}$, $t = 1, \dots, T$, for a range of values for the hyper-parameters and investigate how they affect the mBICV, out-of-sample likelihood, model sparsity, and computation time. The results are shown in \autoref{fig:mbicv}, where thresholded models are in the left and truncated models in the right column. To separate the effect of the selection criterion and  hyper-parameters, \autoref{fig:bic} shows the same results when the BIC is used for selection.

\subsubsection*{Optimal hyper-parameters and models}

The mBICV is shown in the upper panel of \autoref{fig:mbicv}. The mBICV is a convex function of the threshold $\theta$ and has a minimum at $\theta \approx 0.05$. The value $\theta_{\mathrm{auto}} \approx 0.043$ is the threshold selected by the automatic algorithm proposed in \autoref{sec:auto} and is shown as vertical dotted line. The second row shows the out-of-sample log-likelihood obtained based on 5-fold cross-validation. It has a peak at $\theta = 0.45$, which is close the mBICV-optimal parameter. This suggests that the mBICV finds just the right balance between model complexity and fit. Similarly, the convex function of the truncation level with a minimum at $M = 25$. The truncation level selected by the automatic procedure is $M_{\mathrm{auto}} = 26$. The cross-validated log-likelihood has a maximum at $M = 28$, which is only slightly larger.

\autoref{fig:bic} shows the same graphs but using BIC as a selection criterion. We observe that the BIC-optimal models (and the ones selected automatically)  are sub-optimal in terms of cross-validated log-likelihood which confirms that the BIC is inappropriate for high-dimensional models.

\subsubsection*{Level of sparsity}

The sparsity of a vine copula model is characterized by the number of independence copulas. The third rows in \autoref{fig:mbicv} and \autoref{fig:bic} shows the proportion of non-independence copulas  in the model ($q / \kmax$) as function of the hyper-parameters. The proportion of non-independence copulas decreases with the threshold and increases with the truncation level. For both thresholded and truncated models, the mBICV selects models with fewer non-independence copulas (approximately 20\%) compared to BIC (approximately 24\%). The proportions are quite close comparing the thresholded and truncated models, but the truncated model is slightly larger. This again relates to the fact that thresholded models generalize truncated ones.

\subsubsection*{Computation time}

As explained in \autoref{sec:sparse-vines}, sparse vine copulas have a side benefit in terms of computation time, because a large number of pair-copulas are never estimated (but directly set to independence).  We can see this effect in the lower panels of \autoref{fig:mbicv} and \autoref{fig:bic}, where the hyper-parameters are plotted against the time required to fit the model.\footnote{All times were recorded on a single thread of a 8-way Opteron (Dual-Core, 2.6 GHz) CPU with 64GB RAM.}

The larger the threshold, the less time it takes to fit the model. The full model ($\theta = 0$) takes almost two hours, the BIC-optimal thresholded model takes roughly one hour; the mBICV-optimal model takes only 25 minutes. However, there is an overhead for selecting the threshold. Finding the mBICV-optimal threshold and fitting the model takes 35 minutes in total, which is still considerably faster than fitting the full or BIC-optimal models. The difference between the two can be expected to increase with the dimension and level of sparsity. 

Similarly, truncated models with a larger truncation level take more time to fit. The mBICV-optimal truncated model ($M = 26$) and the one selected by the automatically take a bit less than one hour (for truncated models, there is no overhead for selection). This is much less than for the full model ($M = 94$, 1.7 hours) or the BIC-optimal one ($M = 88$, 1.7 hours). But it is twice the time it took to fit the thresholded model. This can be explained by the fact that even in in the first 25 trees there are many independence copulas that were ruled out in the optimal thresholded model. 

We conclude that despite the overhead of threshold selection, the mBICV-optimal thresholded model is much faster to fit than BIC-optimal or truncated models.

\subsection{Out-of-sample Value-at-Risk forecasts}

\begin{figure}
\centering
\includegraphics[width = 0.6\textwidth]{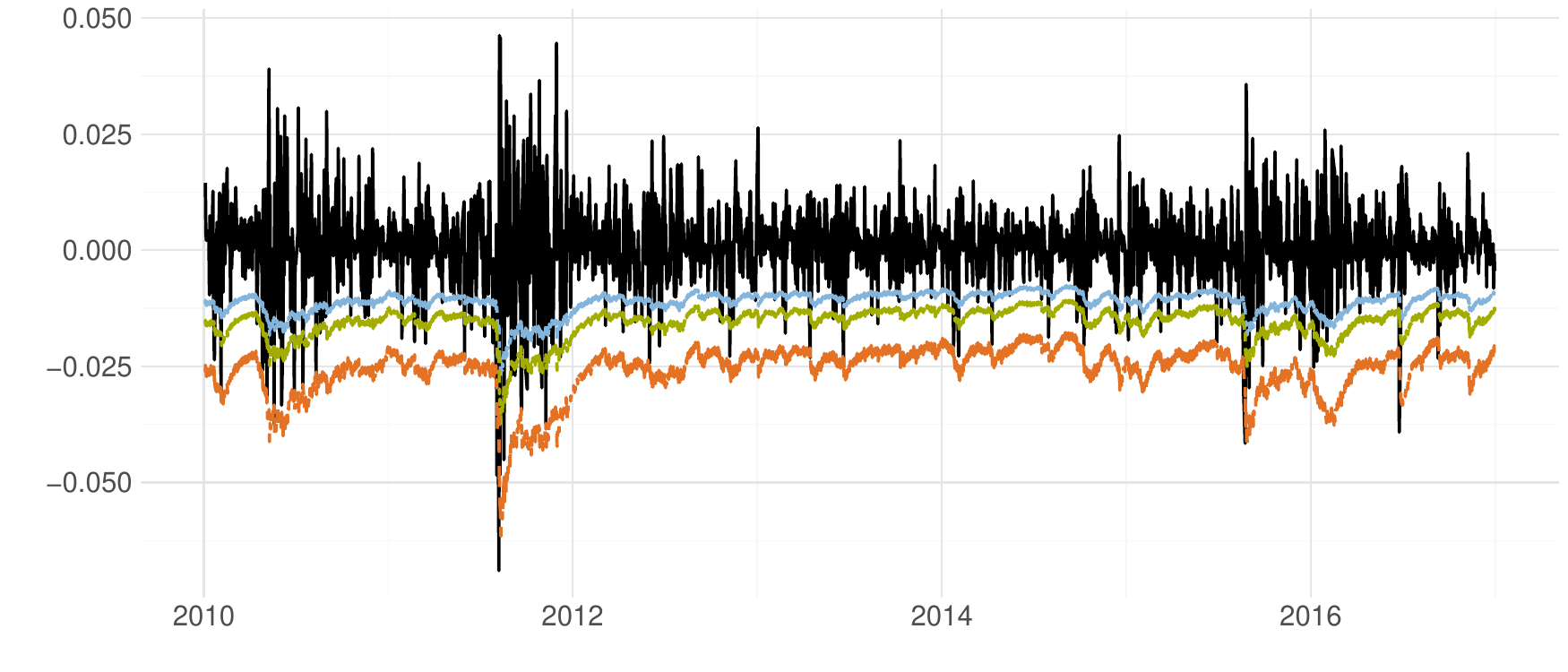}
\caption{Time series of an equally weighted portfolio of S\&P 100 stocks, and VaR predictions at the 90\%, 95\%, and 99\%  levels.}
\label{fig:ts}
\end{figure}

We now analyze the models' ability to accurately forecast the Value-at-Risk out-of-sample. We fit the marginal and vine copula models on a training period period $t = t^* - T_{\mathrm{train}}, \dots, t^*$, and predict the Value-at-Risk for subsequent days $t = t^* + 1, \dots, t^* + T_{\mathrm{test}}$. To ensure that a reasonable amount of data is available, we choose $T_\mathrm{train} = 1260$ (five years) and $T_\mathrm{test} = 252$ (one year). After each year, the marginal and vine copula models are fit again to the previous five years of data. To make the results comparable to the in-sample analysis, we want to use the same period for evaluating out-of-sample forecasts. To achieve this, we augment the data by five more years of data for the period 2005-2009. Four more stocks have to be dropped, leaving us with $d = 92$.

The fitted models can be used to forecast the one-day-ahead Value-at-Risk (VaR) of an equally weighted portfolio of all $d = 96$ stocks. The portfolio return at time $t$ is then $y_t = d^{-1} \sum_{k = 1}^d  x_{t, k}$. The theoretical $\alpha$-level portfolio VaR on day $t+1$ is defined as the $(1 - \alpha)$-quantile of $y_{t +1}$,
\begin{align*}
\mathrm{VaR}_{t+1, \alpha} = \inf\{y \in \R\colon P(y_{t +1} \le y) \ge 1 - \alpha \}, \quad \alpha \in (0, 1).
\end{align*}

Our goal is to forecast $\mathrm{VaR}_{t + 1, \alpha}$,  based on fitted marginal and vine copula models, and only using past observations of the processes $\wh \epsilon_t$ and $\wh \sigma_t$. We proceed as follows:
\begin{enumerate}[1.]
\item Simulate $\bm u_{t+1}^{(r)}$, $r = 1, \dots, R = 10^6$, from the fitted vine copula model. 
\item For all $k = 1, \dots, d$, $r = 1, \dots, R$, set  
\begin{align*}
\epsilon_{t+1, k}^{(r)} &= \Psi^{-1}(u_{t+1}^{(r)}; \wh \nu_k),\\ 
\wh \sigma_{t+1, k} &= \wh \omega_k  + \wh \beta_k \wh \sigma_{t, k}^2 + \wh \alpha_k \wh \epsilon_{t, k}^2, \\
x_{t +1, k}^r &=  \wh \mu_k  + \wh \phi_k x_{t, k} + \wh \psi_k \wh a_{t, k}  +  \wh \sigma_{t + 1, k} \epsilon_{t + 1, k}^{(r)}, \\ 
y_{t+1}^r &= \frac 1 d \sum_{k = 1}^d x_{t+1, k}^r.
\end{align*}
\item The forecast $\wh{\mathrm{VaR}}_{t + 1, \alpha}$ is the empirical $\alpha$-quantile of $(y_{t+1}^r)_{r = 1, \dots, R}$.
\end{enumerate}
Note that although the models are fitted only on the returns in the training period $\{t^* - T_{\mathrm{train}}, \dots, t^*\}$, also all observations in the period $\{t^* + 1, \dots, t \}$ enter the one-day-ahead prediction at time $t+1$ through the residual $\widehat \epsilon_t$. 

\begin{table}[t]
\centering

\begin{tabular}{r|lll|lll|lll}
  \multicolumn{1}{c}{} & \multicolumn{3}{c}{BIC} & \multicolumn{3}{c}{$\theta_{\mathrm{auto}}$} & \multicolumn{3}{c}{$M_{\mathrm{auto}}$}  \\
  $\alpha$ &  0.9 & 0.95 & 0.99 & 0.9 & 0.95 & 0.99 & 0.9 & 0.95 & 0.99 \\ 
  \hline
 exceedances &   0.099 & 0.058 & 0.015 & 0.099 & 0.058 & 0.012 & 0.104 & 0.061 & 0.016 \\ 
 $p$-value &   0.644 & 0.149 & 0.120 & 0.605 & 0.135 & 0.055 & 0.653 & 0.082 & 0.056
\end{tabular}
\caption{Exceedance frequencies of out-of-sample Value-at-Risk forecast and $p$-values of the conditional coverage test of \citet{christoffersen1998} for three models: the one selected by BIC, and thresholded and truncated models selected by the procedure described in \autoref{sec:auto}.}
\label{tab:oos}
\end{table}

\autoref{fig:ts} shows the observed time series $y_t$ along with VaR predictions from an exemplary model. In the following we focus on three models: thresholded and truncated models selected by the automatic procedure from \autoref{sec:sparse-vines} ($\theta_{\mathrm{auto}}$ and $M_{\mathrm{auto}}$), and the non-sparse model selected by BIC as a baseline. We consider three levels for the Value-at-Risk, $\alpha = 0.9, 0.95, 0.99$. The second row of \autoref{tab:oos} shows the frequencies with which the Value-at-Risk forecasts were exceeded. For an optimal model we expect these frequencies to be $1 - \alpha$, which (along with independence of exceedance indicators) is the null hypothesis of  the conditional coverage test  \citep{christoffersen1998}. Corresponding $p$-values are listed in the third row. 

We observe that all models show exceedance frequencies close to our expectation. This is also reflected in the $p$-values, which indicate that no model can be rejected. However, the two sparse models are not only less complex, they were also substantially faster to fit. These two benefits would become even more striking for portfolios of larger size.

    \section{Conclusion}

This article was concerned with model selection in high-dimensional vine copula models. We proposed the mBICV as a selection criterion tailored to sparse vine copula models. It can be used to sequentially select individual pair-copula or automatically select hyper-parameters in sparse model classes. The benefits of the mBICV were illustrated by a case study modeling the dependence in a large stock portfolio. The mBICV-optimal sparse models were shown to produce valid out-of-sample  forecasts for the Value-at-Risk and to be computationally more efficient than models selected by BIC.

We took a first step towards ``large $n$, diverging $d$'' asymptotics in vine copula models by arguing that the the selection criterion is consistent when $d = o(\sqrt{n})$. This claim is supported by simulations,  but our  arguments are only informal due to a lack of results on the behavior of parameter estimates as $d \to \infty$. A formal study of such properties is an interesting direction for future research.

\subsection*{Acknowledgements}
The first and third author were partially supported by the German Research Foundation (DFG grants CZ  86/5-1 and CZ 86/4-1). The authors gratefully acknowledge the compute and data resources provided by the Leibniz Supercomputing Centre (www.lrz.de). The authors are especially grateful to two anonymous referees for many valuable comments and ideas.

    \appendix   
\section{Approximation of error rates}   \label{sec:proof1}

\newcommand{\hetae}{\wh{\eta}_e}
\newcommand{\etaDe}{\bm \eta_{D_e}}
\newcommand{\hetaDe}{\wh{\bm \eta}_{D_e}}
\newcommand{\etaDet}{\eta_{D_{e, t}}}
\newcommand{\hetaDet}{\wh \eta_{D_{e, t}}}
\newcommand{\dmax}{{\overline\delta}}
\newcommand{\peq}{\phantom{=}\;\;}

\subsection{Preliminaries}

Since we consider an edge $e \in E_m$ for a fixed tree level $m < \infty$, we will argue as if the dimension $d$ were fixed. In this case, the usual asymptotic properties of the sequential MLE were established in \citet{hobaekhaff2013} for cases where the margins are estimated parametrically or based on ranks. In particular the estimator is asymptotically normal and satisfies $\| \etah - \etab \|_2 = O_p(n^{-1/2})$. This rate of convergence will only be used for type II error rates, while $o_p(1)$ is sufficient for type I error rates.
We further use the following tail bound for the normal distribution
\begin{align} \label{eq:normal_tail}
\Pr(Z > x) \le \frac 1 {\sqrt{2\pi}} \frac 1 x \exp\biggl(\frac{-x^2}{2}\biggr),
\end{align}
where $Z \sim \mathcal{N}(0, 1)$.

The independence model, $c_{j_e, k_e; D_e}^{0} \equiv 1$, yields
\begin{align*}
\mathrm{mBICV}_e\bigl(0, \etah_{D_e}\bigr) = -2 \ln (1 - \psi_0^m).
\end{align*}
Therefore, the alternative model $c_{j_e, k_e; D_e}^{\etahe}$ is selected if 
\begin{align*}
2 \ell_e\bigl(\etahe, \etah_{D_e}\bigl) > \wh \nu_e \ln n - 2\ln \psi_0^m + 2\ln (1 - \psi_0^m).
\end{align*}
\subsection{Type I error rates}

We start by approximating the type I error $\alpha_{n, e}$.
Let $\hetae$ be the sequential MLE and $\etae = 0$. It holds
\begin{align*}
\ell_e(\hetae, \hetaDe) &= \ell_e(\etae, \hetaDe) + \partial_{\eta_e}\ell_e(\hetae, \hetaDe)(\hetae - \etae)  + \partial_{\eta_e}^2\ell_e(\etae^*, \hetaDe)  (\hetae - \etae)^2,
\end{align*}
where $\eta_e^*$ lies between $0$ and $\hetae$.
The first term on the right is zero because $c_{j_e, k_e; D_e}^{0} \equiv 1$ for $\eta_e = 0$. The second term is zero by the definition of the sequential MLE. Hence,
\begin{align*}
\ell_e(\hetae, \hetaDe) = \partial_{\eta_e}^2\ell_e(\etae^*, \hetaDe)  (\hetae - \etae)^2 = E\{\partial_{\eta_e}^2\ell_e(\etae, \etaDe)\}  (\hetae - \etae)^2\{1 + o_p(1)\},
\end{align*}
where the second equality follows from the law of large numbers and the consistency of $\etah$. 
Now  $E\{\partial_{\eta_e}^2\ell_e(0, \etaDe)\}^{1/2}(\hetae - \etae)$ converges in distribution to a standard normal variable, yielding
\begin{align*}
 \alpha_{n, e}&\approx 2\Pr\bigl(Z > \sqrt{\wh \nu_e \ln n - 2\ln \psi_0^m + 2\ln (1 - \psi_0^m)}\bigr) ,
\end{align*}
where $Z \sim \mathcal{N}(0, 1)$. 
Using \eqref{eq:normal_tail} together with the fact that $\psi_0^m$ lies in $(0, 1)$ and is decreasing in $m$  yields
\begin{align*}
\alpha_{n, e} \approx \frac{2}{\sqrt{2\pi n}} \frac{\psi_0^m(1 - \psi_0^m)^{-1}}{\sqrt{\ln n - 2  \ln \psi_0^m + 2\ln (1 - \psi_0^m)}}  
&\le \frac{2}{\sqrt{2\pi n}} \frac{\psi_0^m(1 - \psi_0)^{-1}}{\sqrt{\ln n + 2\ln (1 - \psi_0)}} 
= O\biggl( \frac{\psi_0^m}{\sqrt{n \ln n}} \biggr).
\end{align*}

\subsection{Type II error rates}

For the type II error, we have
\begin{align*}
\beta_{n, e} &=  \Pr \bigl\{2 \ell_e\bigl(\etahe, \hetaDe\bigr) < \wh \nu_e \ln n - 2\ln \psi_0^m + 2\ln (1 - \psi_0^m) \mid \eta_e \neq 0 \bigr\}.
\end{align*}
We again begin with a Taylor expansion:
\begin{align*}
\ell_e(\hetae, \hetaDe) &= \ell_e(\etae, \etaDe) + \nabla_{\etab}\ell_e(\etae, \etaDe)(\etah - \etab)  + (\etah - \etab)^\top\nabla_{\etab}^2\ell_e(\etae^*, \bm \eta_{D_e}^*)(\etah - \etab).
\end{align*}
Since $E\{\nabla_{\etab}\ell_e(\etae, \etaDe)\} = 0$ by definition, it holds $n^{-1}\nabla_{\etab}\ell_e(\etae, \etaDe) = O_p(n^{-1/2})$ by the law of large numbers. Further, the quadratic term is of order $O_p(1)$ given that $\| \etah - \etab \|_2 = O_p(n^{-1/2})$, whence
\begin{align*}
\ell_e(\wh \etae, \hetaDe) = \ell_e(\etae, \etaDe) + O_p(1).
\end{align*}
We apply the central limit theorem and Slutsky's lemma to the right hand side, yielding 
\begin{align*}
 \frac 1 {\sqrt{n}} \bigr\{\ell_e\bigl(\etahe, \hetaDe\bigl) - \sqrt{n}\mathrm{MI}_e \} \stackrel{d}{\to} \mathcal{N}(0, \sigma_e^2).
\end{align*}
Noting
\begin{align*}
\sqrt{n} \mathrm{MI}_e  - \frac {\wh \nu_e \ln n - 2\ln \psi_0^m + 2\ln (1 - \psi_0^m)}{2 \sqrt{n} \sigma_e} = \sqrt{n} \mathrm{MI}_e\{1 + o(1)\},
\end{align*}
and an application of the tail bound \eqref{eq:normal_tail} suggests
\begin{align*}
\beta_{n, e} &\approx \Pr\bigl(Z > -\sqrt{n} \mathrm{MI}_e\bigr) =  O\biggl(\frac{e^{-nMI_e^2}}{\sqrt{n}MI_e}\biggr).
\end{align*}

	\bibliography{main}
\end{document}